\documentclass[fleqn,usenatbib]{mnras}

\usepackage[T1]{fontenc}

\DeclareRobustCommand{\VAN}[3]{#2}
\let\VANthebibliography\thebibliography
\def\thebibliography{\DeclareRobustCommand{\VAN}[3]{##3}\VANthebibliography}

\usepackage{graphicx}	
\usepackage{amsmath}	
\usepackage{amssymb}	
\usepackage{newtxtext,newtxmath}
\usepackage{mathtools}
\usepackage{lscape}
\usepackage[T1]{fontenc}
\usepackage[utf8]{inputenc}
\usepackage{enumerate}
\usepackage{booktabs}
\usepackage[x11names,dvipsnames]{xcolor}
\usepackage{graphicx}
\usepackage{bm}
\usepackage{afterpage}
\usepackage{color}
\usepackage{array}
\usepackage{amsfonts}
\usepackage{floatflt}
\usepackage{attachfile}
\usepackage{xurl}
\usepackage{fancyhdr}
\usepackage{multicol}
\usepackage{caption}
\usepackage{booktabs,array}
\usepackage{multimedia}
\usepackage{geometry}
\usepackage{lipsum}
\usepackage{tikz}
\usepackage{wrapfig}
\usepackage{floatrow}
\usepackage{gensymb}
\usepackage[framemethod=Tikz]{mdframed}
\usepackage{dcolumn}
\usepackage{bm}
\usepackage[font=small,skip=-60pt]{caption}
\usepackage{setspace}
\usepackage{subcaption}
\usepackage{hyperref}
\hypersetup{
    colorlinks=true,
    linkcolor=black,
    filecolor=magenta,      
    urlcolor=black,
}

\title[Testing sonification of light curves with {\tt astronify}]{Evaluating the efficacy of sonification for signal detection in univariate, evenly sampled light curves using {\tt astronify} }

\author[J. Tucker Brown et al.]{
J. Tucker Brown,$^{1}$\thanks{E-mail: jacktuckerbrown@gmail.com}
C.M. Harrison,$^{1}$\thanks{E-mail: christopher.harrison@newcastle.ac.uk}
A. Zanella$^{2}$ and
J. Trayford$^{3}$
\\
$^{1}$School of Mathematics, Statistics and Physics, Newcastle University, NE1 7RU, UK\\
$^{2}$Istituto Nazionale di Astrofisica, Vicolo dell’Osservatorio 5, 35122, Padova, Italy\\
$^{3}$Institute of Cosmology and Gravitation, University of Portsmouth, Dennis Sciama Building, Burnaby Road, Portsmouth PO1 3FX, UK
}

\pubyear{2021}

\begin{document}
\label{firstpage}
\pagerange{\pageref{firstpage}--\pageref{lastpage}}
\maketitle

\begin{abstract}
Sonification is the technique of representing data with sound, with potential applications in astronomy research for aiding discovery and accessibility. Several astronomy-focused sonification tools have been developed; however, efficacy testing is extremely limited. We performed testing of {\tt astronify}, a prototype tool for sonification functionality within the Barbara A. Mikulski Archive for Space Telescopes (MAST). We created synthetic light curves containing zero, one, or two transit-like signals with a range of signal-to-noise ratios (SNRs=3--100) and applied the default mapping of brightness to pitch. We performed remote testing, asking participants to count signals when presented with light curves as a sonification, visual plot, or combination of both. We obtained 192 responses, of which 118 self-classified as experts in astronomy and data analysis. For high SNRs ($=$30 and 100), experts and non-experts performed well with sonified data (85--100\% successful signal counting). At low SNRs ($=$3 and 5) both groups were consistent with guessing with sonifications. At medium SNRs ($=$7 and 10), experts performed no better than non-experts with sonifications but significantly better (factor of $\sim$2–3) with visuals. We infer that sonification training, like that experienced by experts for {\em visual} data inspection, will be important if this sonification method is to be useful for moderate SNR signal detection within astronomical archives and broader research. Nonetheless, we show that even a very simple, and non-optimised, sonification approach allows users to identify high SNR signals. A more optimised approach, for which we present ideas, would likely yield higher success for lower SNR signals.   
\end{abstract}




\section{Introduction}
\label{sec:intro}
There is a growing interest within the astronomical community to use sound to represent data or concepts \citep[][]{Harrison2022Nat,Zanella2022}. This is evidenced through the 104 astronomy entries that have been submitted to the Data Sonification Archive, which is a community-driven effort to collate and record projects that have used sound to represent data\footnote{Information correct as of 7th September 2022, collected from searching for the term ``astronomy'' at: \url{https://sonification.design}.}. These projects are designed for a range of applications. This includes using sound design for astronomy public engagement and education, which often have a primary or a secondary goal to increase accessibility to blind and low vision audiences \citep[e.g.,][]{Bieryla2020,Harrison2022AG,Bardelli2022,GarciaBenito2022}. There are also a growing number of tools using sound to represent astronomical and space science data, i.e., through the process of `sonification' \citep[][]{Kramer1999}, to aid scientific discovery in astronomy and space sciences and to simultaneously increase accessibility \citep[e.g.,][]{DiazMerced2008,DiazMerced2013,Archer2018,Cooke2019,STRAUSS,Archer2022}.

The tools being developed aim to utilise the huge potential of using sound for data exploration and discovery. Sound is inherently multi-dimensional (e.g., pitch, volume, duration, timbre), we can perceive several sound streams simultaneously and we are effective at identifying recognisable sounds amidst noise  \citep[][]{Hermann2011,Sawe2020}. Therefore, there is an impressive theoretical potential in using sound to intuitively and efficiently explore complex and/or multi-dimensional datasets, such as those increasingly used for astronomy (\citealt{Cooke2019,Sawe2020}). Furthermore, sound can be more optimal than visuals for inspecting time-based information, such as transient events, and sound does not necessarily require our focused attention to monitor the data \cite[e.g.,][]{Shams2000,Guttman2005}. Indeed, \cite{Cooke2019} hope to take advantage of these benefits, by using their sonification tool {\tt StarSound}\footnote{\url{https://www.jeffreyhannam.com/starsound}}, to enhance and accelerate scientific discovery as part of the transient monitoring and multi-messenger program `Deeper Wider Faster' \citep{Andreoni2019}. A general review on sonification benefits and applications is provided by \cite{Hermann2011} and \cite{Zanella2022} review applications of sonification for astronomy education, public engagement, and research. 

There are a handful of examples of where sonification helped make discoveries within astronomy and space science data \citep[see discussion in][]{DiazMerced2013,Alexander2014,Zanella2022}. For example, data sonification revealed a characteristic sound pattern that helped lead to the discovery of a problem during the Voyager 2 space mission caused by high-speed collisions with charged micrometeoroids \citep[][]{Scarf1982}. Sonification also helped identify the most promising carbon ionic ratio to measure the solar wind type \citep[][]{Alexander2011,Landi2012}. Exploring the use of sound also naturally opens up the possibility to make research as a career more accessible to the blind and low vision community, assuming that the tools themselves are also accessible \citep[][]{Garcia2019,Casado2021,NoelStorr2022}.

Despite the promising potential of sonification for astronomical research, it has not yet been widely adopted. There are many challenges to consider, including developing standardised and optimised approaches for turning the data into sound \citep{Harrison2022Nat,Zanella2022}. Here we focus on one particular challenge, which is the lack of efficacy testing of the sonification methods and tools that are being developed for astronomical data exploration. Proof that the sonification methods and tools being developed are effective in achieving their goals will be crucial for wider adoption. To our knowledge, there are no published studies of extensive user testing of the latest astronomy sonification tools \cite[although see][for small focus group testing on earlier methods]{DiazMerced2013}. 

With this study we perform the first efficacy testing of a new sonification tool, {\tt astronify}\footnote{\url{https://astronify.readthedocs.io/}}, using voluntary participants. {\tt astronify} is a {\tt Python} package produced by a team at the Space Telescope Science Institute, and the developers aim to develop and incorporate this sonification tool into one of the world's biggest astronomical archives: the Barbara A. Mikulski Archive for Space Telescopes (MAST)\footnote{\url{https://archive.stsci.edu/}}. Adding sonification functionality to such a widely used archive would be a major step towards making sonification more of a mainstream approach to inspect astronomical data and increasing accessibility. Although currently the functionality of {\tt astronify} is only built for one dimensional data series (e.g., light curve data and spectra), the creators state that the package ``will ultimately grow to encompass a range of sonification functionality''\footnote{Quote taken from website, \url{https://astronify.readthedocs.io/en/latest/}, June 2022.}, which may include images and other multi-dimensional data. Therefore, this tool was chosen as appropriate for testing the efficacy of the sonification approach within the context of astronomical data. For these first tests, we chose to use the default sonification approach of  {\tt astronify} and we designed a simple experiment to establish how effectively participants, both with expert and non-expert knowledge, could identify signals in the sonified data across a range of signal-to-noise ratios (SNRs).

In Section~\ref{sec:LightCurves} we describe how we used {\tt astronify} to produce synthetic  light curves and the corresponding sonification. In Section~\ref{sec:Testing} we describe our testing through online surveys completed by voluntary participants. In Section~\ref{sec:Results} we present our results before discussing them in Section~\ref{sec:Discussion}. Finally  we present our conclusions in Section~\ref{sec:Conclusions}.

\section{Production of synthetic  data with {\tt astronify}}
\label{sec:LightCurves}

Our experiment was to ask participants to count how many signals that they could identify in synthetic light curve data when presented with them as a sonification, a visual plot, or, a combination of these two. Our goals were to compare the participant's success rates of correctly counting the signals between these different data formats as a function of SNR. We used the {\tt astronify} package to both produce the synthetic light curves and to produce the sonification of these data. 

We opted for synthetic light curves, as opposed to real data, to ensure that we had full control of what was in the data that we asked the participants to inspect. Furthermore, for this initial study, we chose for a simple approach of injecting transit-like features into univariate light curves, which is easily supported by the current functionality of {\tt astronify} (described below). These signals corresponds to periodic drops in brightness and such data could be applicable to exoplanet transit data and eclipsing X-ray binary data, as well as data of other astronomical transient and periodic phenomena. 

In this initial study we did not aim to produce a synthetic dataset that would be representative of the variety of real transit light curve data (this would be a natural next step; see discussion in Section~\ref{sec:limitations}). Instead, our goal was to make these data and signals as simple as possible, and suitable for a controlled experiment with voluntary participants. Therefore, we varied as few parameters as possible of the `signals' within a noisy background and ensured that multiple signals could be presented to the participants in a reasonable amount of time during a sonification (i.e., of the order of seconds). 

\begin{figure*}
    \includegraphics[width=0.99\textwidth]{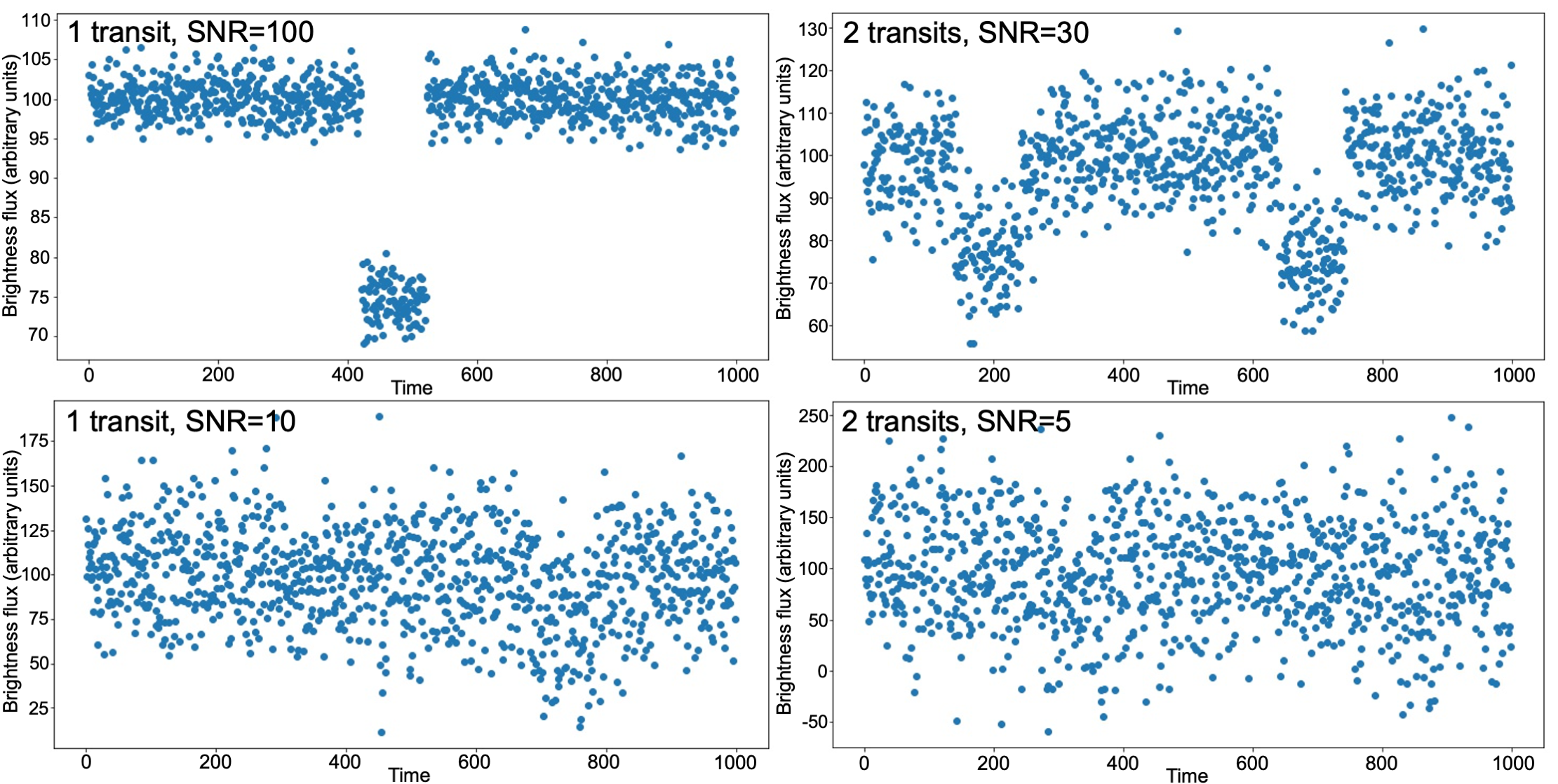}
    \caption{Example synthetic light curves that were used in the participant testing. Time is in arbitrary units, but covers 0--10 seconds in the sonification version of the data. {\em Top-left:} One transit with SNR$=$100. The sonification of this light curve is available through a link: \href{https://data.ncl.ac.uk/articles/dataset/Tucker-Brown_et_al_-_Astronify_Efficacy_Testing_-_Data/20936749?file=37222321}{here.} {\em Top-right:} Two transits with SNR$=$30. Sonification available: \href{https://data.ncl.ac.uk/articles/dataset/Tucker-Brown_et_al_-_Astronify_Efficacy_Testing_-_Data/20936749?file=37222327}{here}. {\em  Bottom-left:} One transit with SNR$=$10. Sonification available: \href{https://data.ncl.ac.uk/articles/dataset/Tucker-Brown_et_al_-_Astronify_Efficacy_Testing_-_Data/20936749?file=37222318}{here}. {\em  Bottom-right:} Two transits with SNR$=$5. Sonification available: \href{https://data.ncl.ac.uk/articles/dataset/Tucker-Brown_et_al_-_Astronify_Efficacy_Testing_-_Data/20936749?file=37222324}{here}. The labels of SNR values and transit numbers were not presented to the participants during testing.}
    \label{fig:examples}
\end{figure*}

\subsection{{\tt astronify} and synthetic light curves}
\label{sec:mocks}
{\tt astronify} is a scriptable {\tt Python} package, currently with no graphical user interface, that allows users to read in time series data. In theory, any univariate time series data can be quickly sonified inside the {\tt astronify} package. It is particularly compatible with the {\tt lightkurve} package \citep[][]{LightKurve}, which enables light curve data taken with the {\em Kepler} (\citealt{Borucki2010}) and {\em TESS} (\citealt{Ricker2015}) space telescopes to be directly loaded within a script. However, for the reasons described above, we generated {\em synthetic} light curves using {\tt astronify's} light curve simulator.

We used the {\tt simulated\_lc} function with the `transit' option\footnote{The function and the variables are described here: \url{https://astronify.readthedocs.io/en/latest/api/astronify.simulator.simulated_lc.html\#astronify.simulator.simulated_lc}}. This produces synthetic data in the form of brightness as a function of time, with the addition of drops in brightness (the transits), which are boxcar functions with a specified width and depth. To reduce the number of variables, we fixed the total light curve length to $t_{l} = 1000$ (arbitrary units), the fractional transit depths to $t_{d} = 25$ and the transit width as a fraction of the total length to be $w_{f} = 0.1$. We only altered the transit periods, to ensure that zero, one, or two transits would appear within the synthetic light curves. Finally, we used the package's feature to add Gaussian noise, such that the signals had a range of SNRs. SNR was estimated to be $t_{d}{\sqrt{N_t}}$/$\sigma$, where $N_t$ is the number of data points in the transit and $\sigma$ is the standard deviation of a normal distribution of the random noise. We created light curves with SNRs of 3, 5, 7, 10, 30 and 100. This range was chosen to give a sample of synthetic data curves which had low SNRs (SNR=3 and 5), medium SNRs (SNR=7 and 10) and high SNRs (SNR=30 and 100). 

In total we produced a synthetic dataset of 18 light curves. This included light curves with either one or two signals at each of the six SNRs chosen (i.e., 12 light curves containing signals). The final six light curves all contained zero signals. This large number of zero-signal light curves helped us to estimate the random guessing behaviour of the participants (see Section~\ref{Sec:Guessing}).

In Figure~\ref{fig:examples} we present example synthetic light curves that were shown to participants. These very simple visualisations (i.e., with no binning, running averages or other guide-to-the-eye features) were chosen to be fairly comparable to the simple sonification representations of the same data that we also presented to the participants during the testing. The full set of plots that were shown to the participants are available at this link: \url{https://doi.org/10.25405/data.ncl.20936749}.

\subsection{Sonification}
\label{sec:sonification}
{\tt astronify}'s sonification algorithm is to map brightness to pitch, with positive polarity. That means, a decrease in brightness is represented by a decrease in pitch in the sonification. Each data point is converted into a discrete tone, where the pitch is determined by the brightness and the data points are played sequentially in time. We opted for the default linear mapping of brightness to pitch.

Sonification parameters were kept consistent for all synthetic datasets. Data were sonified with each light curve's full brightness range (minimum--maximum) mapped onto a one octave frequency range of (440--880)\,Hz. This is a comfortable range for participants to listen to regardless of playback device (e.g. headphones, laptop speakers) and corresponds to the octave A$_{4}$--A$_{5}$ in  scientific pitch notation (where C$_{4}$ is ``middle C'' on the Western piano keyboard). The spacing between notes and the note lengths were set to be 0.01 and 0.2, respectively, based on the advise of the software developers, to optimise the listener's experience. Example sonifications of the synthetic light curve are available in Figure~\ref{fig:examples}. We also created movie files, where we combined the static visuals with the sonifications for our combined plot+sonification data representations. The full set of audio files for the sonification and the movie files are available at this link: \url{https://doi.org/10.25405/data.ncl.20936749}.

\subsection{Final data representations}
\label{Sec:DataRep}
In total, the 18 separate synthetic light curves (Section~\ref{sec:mocks}) were all presented in three formats: (1) a visual plot only; (2) a sonification only; and (3) a combined plot+sonification. Therefore, this resulted in a total of 54 individual light curve representations that were used for the participant testing.

\section{Participant testing}
\label{sec:Testing}

We used Google Forms to create six separate surveys, each containing an independent, randomly selected, set of nine of the total 54 data representations (Section~\ref{Sec:DataRep}). Therefore, each survey had a set of light curves containing a random range of SNRs and number of transit `signals'. Furthermore, each of the six surveys contained a mixture of sonification-only, visual-only, and plot+sonification data representations.  The six surveys were randomly distributed among volunteer participants. We created this set of six smaller surveys, rather than one large survey, to minimise fatigue of the participants taking part. A transcript for the survey text, the tutorial (see below), and questions that were presented to the participants are available at this link: \url{https://doi.org/10.25405/data.ncl.20936749}.

\subsection{Survey format}
In order to assess demographics of the participants, the surveys contained initial questions asking for a self-assessment of their expertise in both astronomy and data analysis. These questions were included to obtain an understanding of how familiar the users would be with the concept of light curves and transits and if the users would be familiar with data analysis. The assumption here was that experts in these groups would be experienced in analysing data in visual graphical formats. The users were asked to rank their expertise from a scale of 1 (labelled ``High School'') to 4 (labelled ``PhD/academic''). 

We also asked the participants if they identified as having a vision impairment. This was to control for potential challenges in this group interpreting the visual representations of the data and an initial experimental goal to compare the performance of using sonification between this group with the wider population.

After the demographic questions, the surveys contained a brief tutorial, giving examples of each of the three data formats. Each participant experienced the same examples but we avoided using the same synthetic light curves that were included in the main survey questions (see Section~\ref{sec:LightCurves}). These covered a high, low and medium SNR example, enabling the participants to become familiar with what low and high SNR signals would look and sound like.

After the demographic questions and tutorial examples, the surveys contained nine questions, covering a random subset of the 54 data representations. The participants could select 0, 1, 2, 3 or 4 as the number of signals that they identified in the data. We also included a `N/A' option, for example, to account for situations such as the users who were unable to see the visual plots due to sight loss. Options of three and four were included, despite none of the synthetic light curves having these number of signals, in order to help assess the random guessing behaviour of the participants (see Section~\ref{Sec:Guessing}).

\subsection{Survey deployment and demographics of participants}
\label{sec:demographics}
Survey links were distributed to potential volunteer participants via emails and the surveys were open for two separate periods: (1) from 22nd March 2021 to 16th April 2021 and (2) from 25th May 2021 to 12th August 2021. In the instructions participants were advised to use headphones to complete the survey, that there was no time limit placed to complete the survey, the users were free to inspect the data for as long as they wished and they could re-listen to the sonifications if they wished to. We discuss possibilities for more controlled conditions in Section~\ref{sec:limitations}. 

We note that for each of the six surveys, we only received between 1 and 3 responses from participants who identified as having a vision impairment. Therefore, these numbers are unfortunately too low to make any significant conclusions and we exclude these responses from our final analysis in this particular study. We note that to test the application of sonification for increasing accessibility for blind and low vision scientists would likely require a focus group study, inviting members of this community to participate to test both the sonification method itself and also the accessibility of the tools themselves \citep[][]{Garcia2019,Casado2021}. This will be important future work. After removing the responses from these participants, we received a total of 192 survey responses. From these, each of the six surveys were completed between 27 and 38 times each.

\begin{figure}[!ht]
    \includegraphics[width=0.9\linewidth]{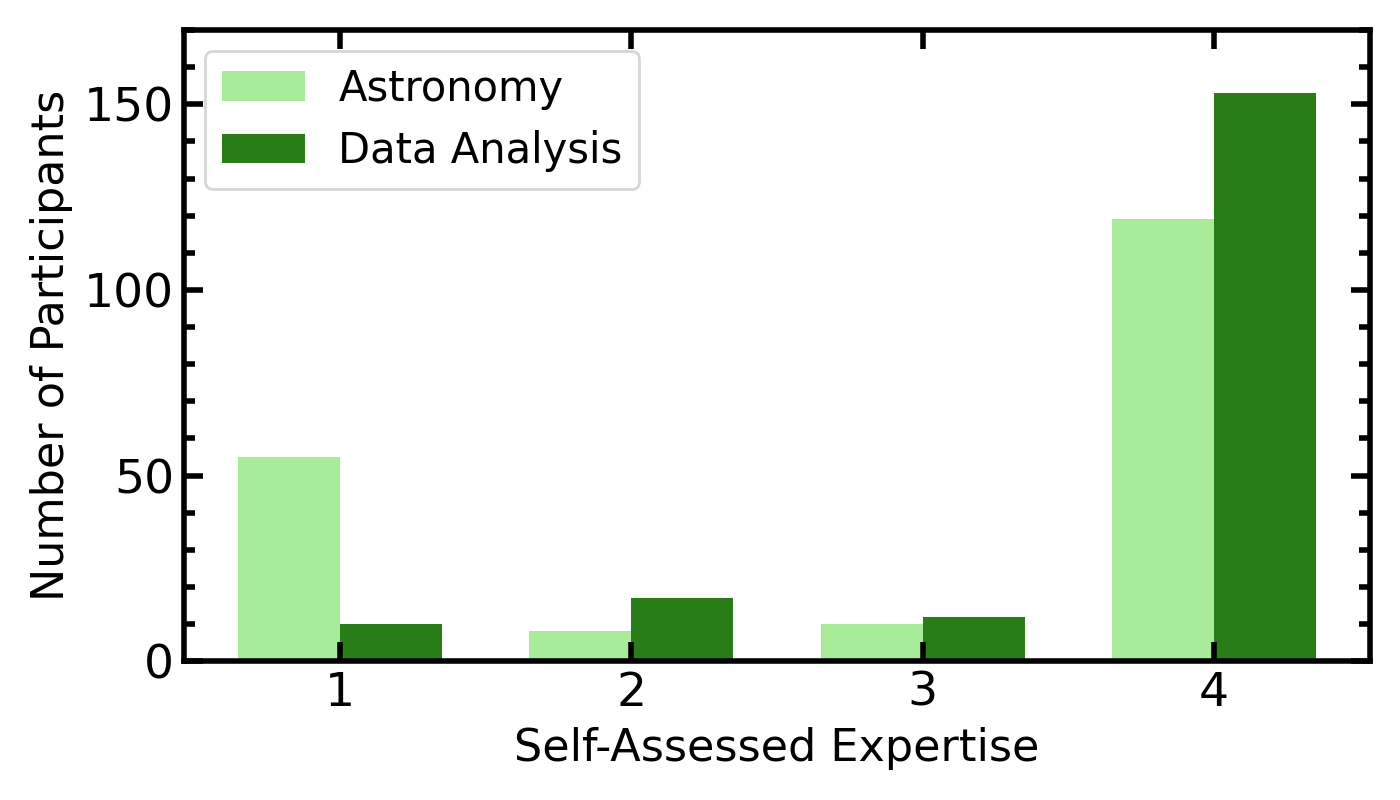}
    \caption{The distribution of responses to the self-assessment questions of expertise using a scale of 1 (low) to 4 (high) in astronomy (lighter green bars) and data analysis (darker green bars). These data are tabulated in Table~\ref{Tab:demographics}.}
    \label{fig:demographics}
\end{figure}

We did not specifically target any groups to complete the surveys; however, the surveys were mostly distributed within the personal and professional networks of the authors. Unsurprisingly, this led to a higher fraction of self-assessed experts in astronomy and data analysis, compared to non experts. For example, 119 responses selected the top level answer of 4 for their expertise in astronomy, compared to 73 responses for all answers of 1, 2 and 3. We take into account the different number of responses across different questions and the demographic groups using appropriate confidence intervals. The full demographic data based on the astronomy and data analysis expertise is visually represented in Figure~\ref{fig:demographics} and is tabulated in Table~\ref{Tab:demographics}.

\begin{table}
\caption{The distribution of participants' self-assessed expertise on a scale of 1 (low) to 4 (high) in astronomy and data analysis.}
\label{Tab:demographics}
\begin{tabular}{rrr}
\toprule
Expertise &  Astronomy &  Data analysis \\
\midrule
                     1 (low) &         55 &             10 \\
                     2 &          8 &             17 \\
                     3 &         10 &             12 \\
                     4 (high) &        119 &            153 \\
\bottomrule
\end{tabular}
\end{table}

We split the participant responses into two main demographic groups: `Experts', who were defined as users who put 4/4 for their expertise in both astronomy and data analysis (118 responses) and `Non Experts' who answered below 4 to {\em both} of these questions (38 responses). Given the significant overlap in the expertise assessment answers for data analysis and astronomy expertise, we did not explore these two categories individually. 

There were also 36 participants who responded with 4/4 for their expertise in either astronomy or data analysis, but not both. We define this sub-group as `Partial Experts'. For completeness we provide the results of this group in Table~\ref{tab:data}. However, for the rest of the study we focus only on the Experts and Non Experts to have a clear distinction in the expertise of the comparison groups. However, we note that the Partial Experts group have consistent results to the Experts group (within errors)\footnote{There are only two exceptions to this: (1) the Partial Experts have a higher success rate than Experts for the SNR=5 plots; and (2) they have a higher guess percentage (see Section~\ref{Sec:Guessing}) for the plots$+$sonifications. Nonetheless, neither of these two exceptions change the overall trends or conclusions of the work.}.

\subsection{Calculation of success rates}
\label{sec:Analysis}

We have defined success rates as the percentage of the responses that correctly identified one or two signals in the data. We did this separately for each of sonifications, plots, and the combined plot+sonification data representations. We calculated these separately for the Expert and Non Expert demographic groups.

To assess the uncertainty on these percentage success rates, we calculated Wilson score binominal confidence intervals, which take into account both the total number of answers and the number 
of correct answers. The quoted uncertainties on percentage success rates are assuming a confidence of interval of 68.3\%.

The number of total responses, the number of correct responses, the resulting percentage success rates and the uncertainties are all provided in Table~\ref{tab:data}. 

\subsection{Guessing and random behaviour} \label{Sec:Guessing}
To assess if the success rates are consistent with, or better than, the participants randomly guessing one or two signals, we calculated a `guess percentage'. We note that we can not simply assume that the participants' random guesses will be uniformly distributed across the possible answers. For example, there was a tendency for the participants to select zero as the number of signals that they identified when there were actually signals in the data but with low SNRs. 

We sought to quantify how often a participant randomly (i.e., incorrectly) selected one or two signals as their answer. To do this, we calculated the number of times an answer of one or two signals was chosen for the synthetic light curves that actually contain zero signals. Again, Wilson score binominal confidence intervals were calculated to provide an uncertainty on these values. We did this separately for each of the sonifications, visual plots, and plots+sonifications data representations. We also calculated these separately for the Expert and Non Expert demographic groups. 

Across the different data formats and the two expertise groups, the guess percentages range from 4--21\%. The Non Expert group tended to randomly select one or two signals more often than the Expert group. For example, the guess percentage is only $\sim$4\% for the Expert group for the two data representations that include visuals. This is caused by the experts' tendency to guess zero signals, as opposed to one or two signals, when they could not identify a signal in the data. Although the difference is only at the $\sim$2$\sigma$ level, it is interesting that the Experts are 3$\times$ more likely to guess with the sonifications only (with a 12\% guess percentage), and at a consistent rate to the Non Experts, compared to when presented with data formats including a visual. We discuss this in the context of a lack of training in using sonification across both groups in Section~\ref{sec:Discussion}.

The guess percentage values are tabulated in Table~\ref{tab:data}. In the following figures, guess percentages are represented as dashed lines. The semi-transparent boxes around these lines represent the confidence intervals on these values. For guidance, if a success rate percentage has a confidence interval that overlaps with the guess percentage confidence interval, we consider it to be consistent with the participants randomly guessing the answers. If a success rate percentage is above the guess percentage and its confidence intervals do not overlap with the guess percentage confidence intervals, we consider this to be better than randomly guessing.

\begin{figure}
      \includegraphics[width=0.98\linewidth]{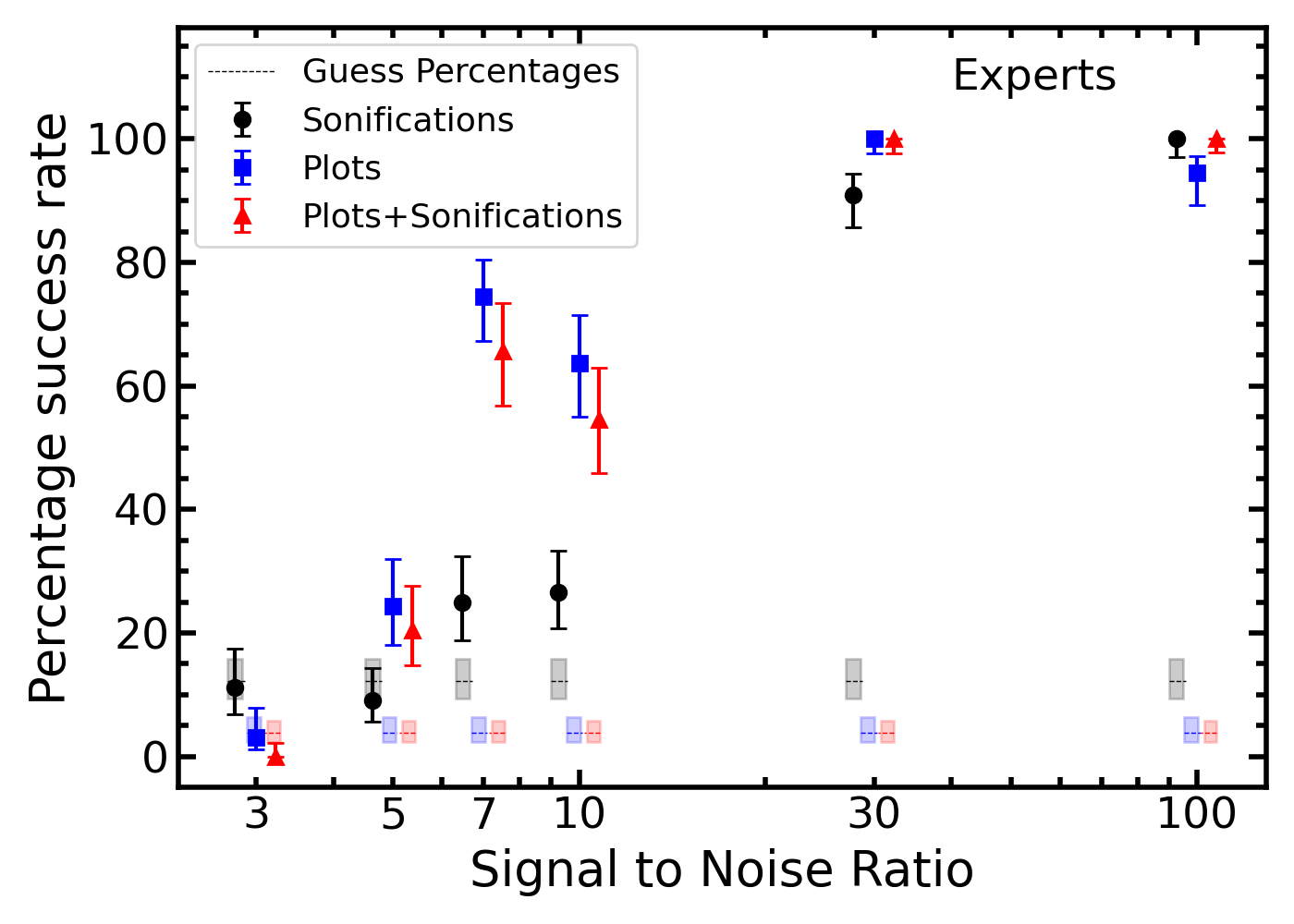}
      \includegraphics[width=0.98\linewidth]{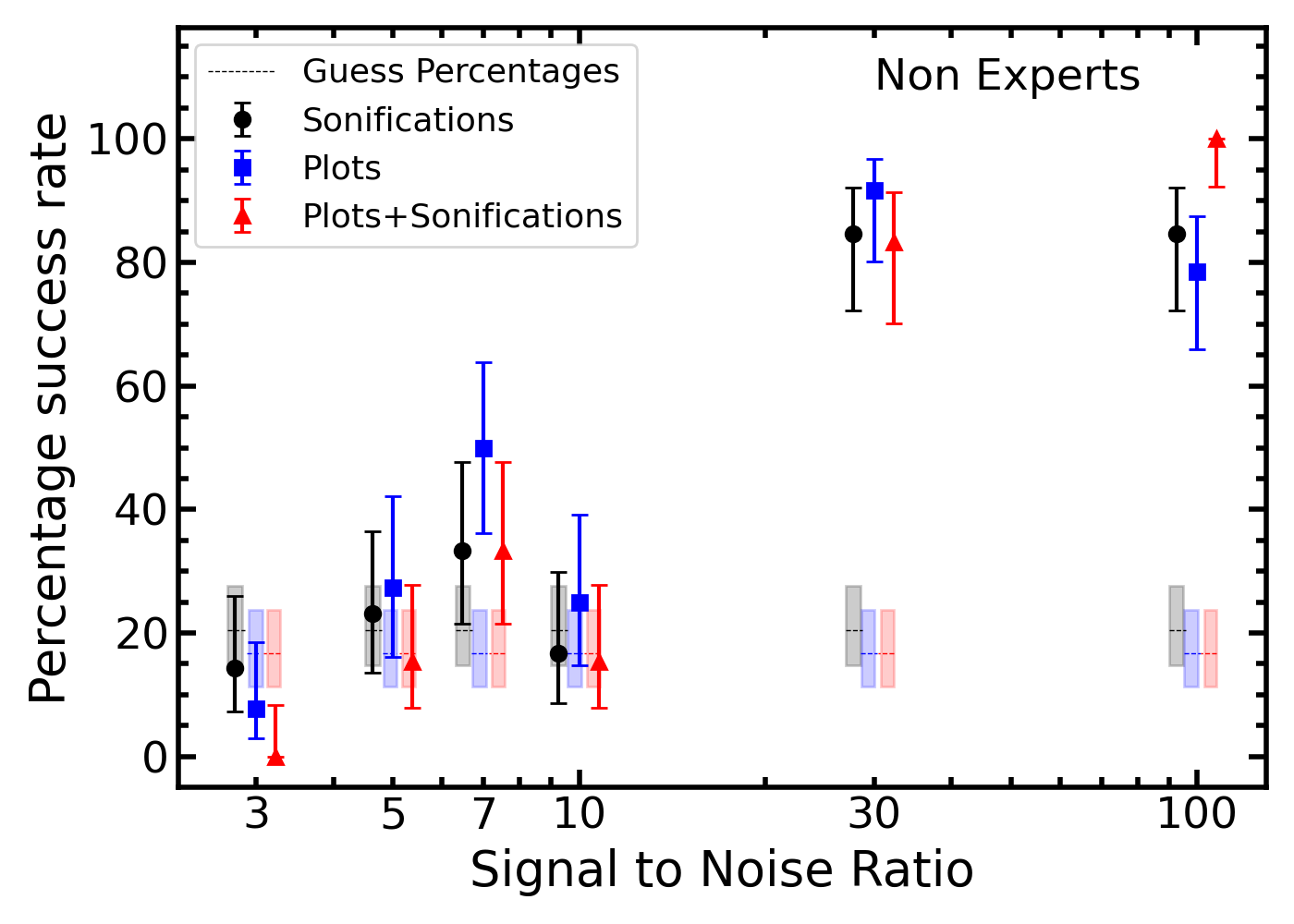}
     \caption{Percentage success rates of correctly identifying the number of signals in the synthetic light curves as a function of SNR. The error bars correspond to the 68.3\% confidence intervals. These success rates are shown for both the Expert group ({\em Top Panel}) and Non Expert group ({\em Bottom Panel}). Success rates are shown separately for data presented as sonifications (circles), visual plots (squares) and combined plots+sonifications (triangles), with a small arbitrary shift in SNRs for clarity of the data points. The estimated success rates for purely random guesses for each data representation are shown as dotted lines. The surrounding shaded regions correspond to the 68.3\% confidence intervals on these guess percentages (see Section~\ref{Sec:Guessing}).}
      \label{fig:successRates}
\end{figure}

\section{Results}
\label{sec:Results} 

\begin{table*}
\caption{Total responses, number of correct answers (identifying the correct number of signals in the data) and the corresponding percentage success rates for the groups of Experts (top), Non Experts (middle), and Partial Experts (bottom). Success rates are given for each of the SNRs trialed, and each of the data formats (sonifications only, plots only, and plots+sonifications). We also display the combined results for medium SNR examples as 7\&10 in the bottom row for each group. The top row for each group provides the estimated percentage consistent with random guesses for each data format (see Section~\ref{Sec:Guessing}).}
\centering{
\begin{tabular}{llllllllll}
\toprule
                       & \multicolumn{3}{c}{Sonifications} & \multicolumn{3}{c}{Plots} & \multicolumn{3}{c}{Plots + Sonifications} \\
                       SNR & Total & Correct & Success rate (\%) & Total & Correct & Success rate (\%) &       Total& Correct & Success rate (\%) \\
\midrule
                   & \multicolumn{9}{c}{Experts (118 participants)}\vspace{0.2cm} \\
       Guess percentage &             -  &              -   &      12$_{-3}^{+4}$ &      -  &             -   &       3.9$_{-1.5}^{+2.4}$&   -  &               - &  3.8$_{-1.4}^{+2.1}$ \vspace{0.05cm} \\
                     3 &              36 &               4 &       11$_{-4}^{+6}$&         32 &               1 &           3$_{-2}^{+5}$&     43 &               0 &  0$_{-0}^{+2}$ \vspace{0.05cm} \\
                     5 &              44 &               4 &        9$_{-3}^{+5}$ &        37 &               9 &          24$_{-6}^{+8}$&     39 &               8 &  21$_{-6}^{+7}$ \vspace{0.05cm} \\
                     7 &              40 &              10 &       25$_{-6}^{+7}$&         43 &              32 &          74$_{-7}^{+6}$&     32 &              21 &  66$_{-9}^{+8}$ \vspace{0.05cm} \\
                    10 &              49 &              13 &       27$_{-6}^{+7}$&         33 &              21 &          64$_{-9}^{+8}$&    33 &              18 &  55$_{-9}^{+8}$ \vspace{0.05cm} \\
                    30 &              44 &              40 &       91$_{-5}^{+3}$&         41 &              41 &         100$_{-2}^{+0}$&     43 &              43 &  100$_{-2}^{+0}$ \vspace{0.05cm} \\
                   100 &              33 &              33 &      100$_{-3}^{+0}$&         36 &              34 &          94$_{-5}^{+3}$&     44 &              44 &  100$_{-2}^{+0}$ \vspace{0.05cm} \\
                   7\&10 &            89 &              23 &       26$_{-4}^{+5}$&         76 &              53 &          70$_{-5}^{+5}$&     65 &              39 &  60$^{+6}_{-6}$ \\
                   &&&&&&&&&\\
                   & \multicolumn{9}{c}{Non Experts (38 participants)} \vspace{0.2cm}\\
Guess percentage       &               - &               -  &        21$_{-6}^{+7}$ &       - &               - &         17$_{-5}^{+7}$ &     -  &             -   &  17$_{-5}^{+7}$  \vspace{0.05cm} \\
                     3 &              14 &               2 &         14$_{-7}^{+12}$ &     13 &               1 &          8$_{-5}^{+11}$&     11 &               0 &   0$_{-0}^{+8}$  \vspace{0.05cm} \\
                     5 &              13 &               3 &        23$_{-10}^{+13}$ &     11 &               3 &        27$_{-11}^{+15}$&     13 &               2 &  15$_{-7}^{+12}$\vspace{0.05cm} \\
                     7 &              12 &               4 &        33$_{-12}^{+14}$ &     12 &               6 &        50$_{-14}^{+14}$&     12 &               4 &  33$_{-12}^{+14}$ \vspace{0.05cm} \\
                    10 &              12 &               2 &        17$_{-8}^{+13}$ &      12 &               3 &        25$_{-10}^{+14}$&     13 &               2 &  15$_{-7}^{+12}$ \vspace{0.05cm} \\
                    30 &              13 &              11 &        85$_{-12}^{+7}$&       12 &              11 &         92$_{-12}^{+5}$&     12 &              10 &  83$_{-13}^{+8}$ \vspace{0.05cm} \\
                   100 &              13 &              11 &        85$_{-12}^{+7}$ &      14 &              11 &         79$_{-13}^{+9}$&     12 &              12 & 100$_{-7}^{+0}$ \vspace{0.05cm}\\
                    7\&10 &          24 &               6 &         25$_{-8}^{+10}$ &      24 &               9 &         38$_{-9}^{+10}$&     25 &               6 &  24$_{-7}^{+9}$ \\
                     &&&&&&&&&\\
                   & \multicolumn{9}{c}{Partial Experts (36 participants)} \vspace{0.2cm}\\
Guess percentage       &               - &               -  &        11$_{-4}^{+6}$ &       - &               - &         4$_{-2}^{+5}$ &     -  &             -   &  14$_{-5}^{+7}$  \vspace{0.05cm} \\
                     3 &              15 &               0 &         0$_{-0}^{+6}$ &      13 &               0 &          0$_{-0}^{+7}$&     12 &               0 &   0$_{-0}^{+8}$  \vspace{0.05cm} \\
                     5 &              14 &               2 &        14$_{-7}^{+12}$ &     11 &               5 &        46$_{-14}^{+15}$&     9 &               3 &  33$_{-13}^{+17}$\vspace{0.05cm} \\
                     7 &              12 &               3 &        25$_{-10}^{+14}$ &     10 &               8 &        80$_{-15}^{+10}$&     12 &               9 &  75$_{-14}^{+10}$ \vspace{0.05cm} \\
                    10 &              15 &               3 &        20$_{-8}^{+12}$ &      10 &               7 &        70$_{-16}^{+12}$&     10 &               5 &  50$_{-15}^{+15}$ \vspace{0.05cm} \\
                    30 &              10 &              10 &        100$_{-9}^{+0}$&       12 &              11 &         92$_{-12}^{+5}$&     10 &              10 & 100$_{-9}^{+0}$ \vspace{0.05cm} \\
                   100 &              8 &               8 &        100$_{-11}^{+0}$ &      14 &              14 &         100$_{-7}^{+0}$&     10 &              10 & 100$_{-9}^{+0}$ \vspace{0.05cm}\\
                    7\&10 &          27 &               8 &         22$_{-7}^{+9}$ &      20 &              15 &         75$_{-11}^{+8}$&     22 &               14 &  64$_{-11}^{+9}$ \\
\bottomrule
\end{tabular}

}
\label{tab:data}
\end{table*}

In Figure~\ref{fig:successRates} we show the percentage success rates of the participants correctly counting the number of signals in the synthetic light curves across the three data formats, as a function of SNR. As described in Section~\ref{sec:LightCurves}, these synthetic light curves consist of univariate time series data of evenly sampled data points. The data contain zero, one, or two signals in the form of a fixed length and depth reduction of flux (i.e., transit-like signals).

Across both groups (Experts in the top panel and Non Experts in the bottom panel) and all data formats (sonifications, plots, and plots+sonifications), we observe the expected broad trend of higher success rates with increasing SNR. The success rates range from very low (i.e., 0--14)\% for SNR=3 to high (i.e., 79--100)\% for SNR=100. In the following we break down the results in to the regimes of low SNRs (SNR=3 and 5), high SNRs (SNR=30 and 100) and medium SNRs (SNR=7 and 10). We discuss the medium SNRs last as these results are the most interesting for differentiating the behaviour between the Expert and Non Expert groups. 

\subsection{Low-SNR regime} \label{lownsr}
Unsurprisingly the success rates, for correctly identifying signals in the synthetic light curves, are very low ($\le$14\%) for SNR=3 across all three data formats and across both expertise groups. Furthermore, in most cases there is no evidence that the participants were performing any better than simply guessing their answers, based on the comparison between the success rates and the estimated guess percentages (Figure~\ref{fig:successRates}). 

At SNR=5, the Non Expert group still performed consistently with simply guessing their answers across all three data formats. The Experts are still only reaching success rates of $\sim$20\%, which is consistent with the success rates of the Non Expert group. However, in contrast, the Expert group performed better than guessing (by a factor of $\sim$5) for both the plots, and plots+sonifications formats. This is driven by their reduced tendency to {\em guess} that there are signals in the data compared to the Non Expert group when inspecting a visual representation (see Section~\ref{Sec:Guessing}). 

\subsection{High-SNR regime} \label{highsnr}
At high values of SNR the success rates exceed the estimated guess percentage significantly, across all three data formats and both expertise groups. Most interesting for this study, is the result that the sonification only format shows high success rates in both Experts (with rates of 91$_{-5}^{+3}$\% and 100$_{-3}^{+0}$\% for SNRs of 30 and 100, respectively) and Non Experts (with rates of 85$_{-12}^{+7}$\% for both SNR=30 and SNR=100). 

These high success rates, for both groups, show that the employed sonification method is an effective tool for identifying high SNR signals in the light curves, mostly irrespective of the previous experience in astronomy and data analysis. This is despite the sonification method in {\tt astronify} being relatively basic and that it has not (yet) been carefully optimised with extensive testing (see discussion in Section~\ref{sec:Discussion}). 

\begin{figure*}
\includegraphics[width=\linewidth]{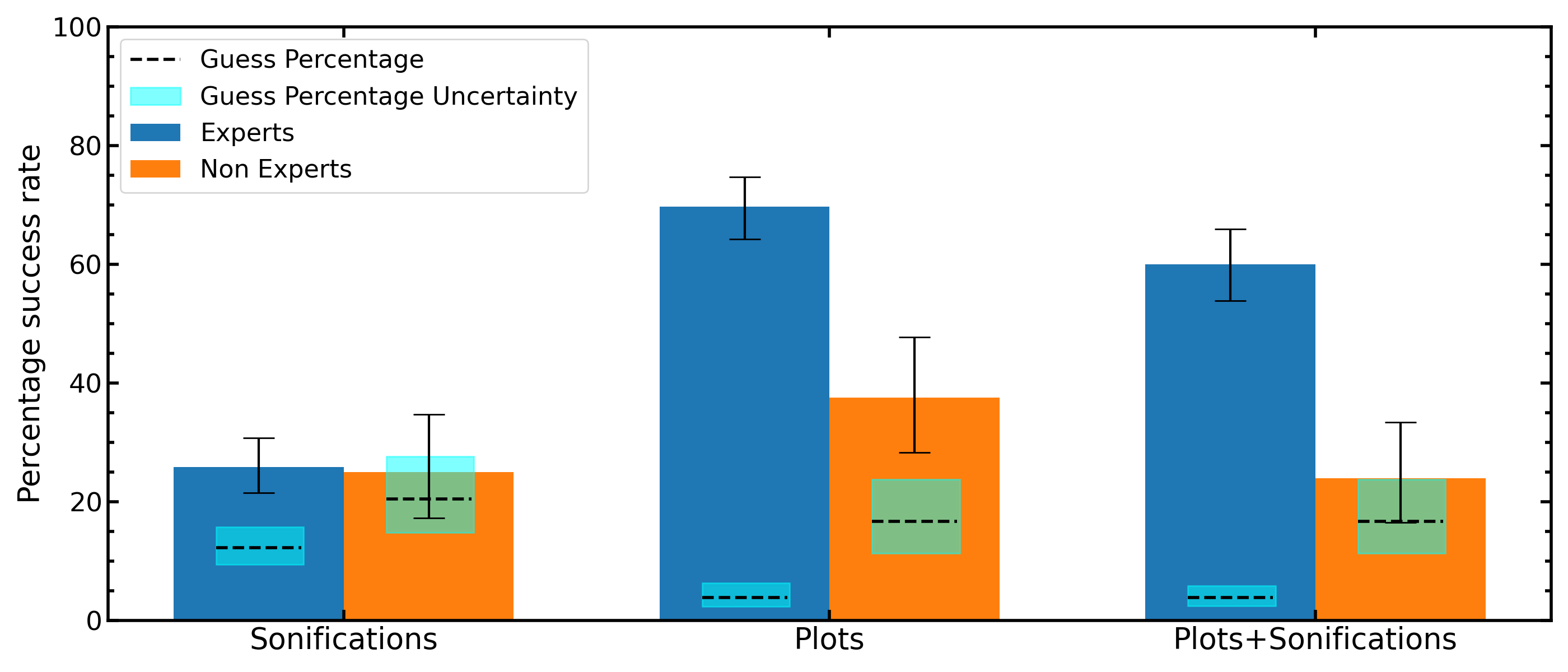}
\caption{Percentage success rates for the medium SNR regime (i.e., combining the results for SNRs of 7 and 10). These are split into the Expert group (blue colour, left bars) and Non Expert group (orange colour, right bars), for each of the three data formats. Error bars show the uncertainties on the guess percentages (see Section~\ref{sec:Analysis}). For each category, dashed lines represent the estimated percentage consistent with random guesses and the surrounding rectangular semi-transparent shaded regions represent the uncertainty on these values (see Section~\ref{Sec:Guessing}).}
\label{fig:snr710}
\end{figure*}

\subsection{Medium-SNR regime} \label{training}
The Expert and Non Expert group differentiate in their results the most in the medium SNR regime. Therefore, to improve the statistics in this regime, we combined the SNR=7 and SNR=10 results and present the success rates visually in Figure~\ref{fig:snr710} (data also tabulated in Table~\ref{tab:data}). However, we note that these combined results provide consistent conclusions to that derived from the individual SNR=7 and SNR=10 results shown in Figure~\ref{fig:successRates}. 

For the sonification format, with the combined SNR=7 and SNR=10 results, the success rates are consistent across the two expertise groups (with success rates of 26$^{+5}_{-4}$\% and 25$^{+10}_{-8}$, for the Experts and Non Experts, respectively). For the Expert group we find that they performed better than randomly guessing with the sonifications, where the guess percentage is 12$^{+4}_{-3}$\%. This is not the case for the Non Expert group; however, this could simply be due to the lower number statistics and/or their slightly higher tendency to guess that they identified one or two signals (Figure~\ref{fig:snr710}). 

For the two formats including a visual representation, i.e., plots only and plot+sonification, the Expert group performed significantly better than they did with the sonification only. The success rates for these formats are 70$^{+5}_{-5}$\% and 60$^{+6}_{-6}$\%, respectively. This is a factor of $\sim$2.5$\times$ higher than the sonification format. In contrast, the Non Expert group did not perform significantly better with these visual representations over the sonification only, and they perform significantly worse than the Expert group (success rates of 38$^{+10}_{-9}$\% and 24$^{+9}_{-7}$\% for the plots only and plots+sonifications, respectively). 

We have shown that in the medium SNR regime, experts and non experts performed equally well at identifying the signals with the sonifications that they inspected. However, the experts perform significantly better when presented with a visual format. As discussed below, we suggest that this is because the experts have had significantly more experience in analysing data visually, whilst both groups have equally little experience using sonifications to inspect data.

\section{Discussion}
\label{sec:Discussion}
The {\tt astronify} package is designed as a proto-type to perform sonification of astronomical datasets, with the long-term goal to add sonification functionality to the MAST archive. However, in order for this to be successful and to be widely utilised by the astronomical community it is important that user testing is carried out to assess the efficacy of the sonification approaches taken (see discussion in \citealt{Harrison2022Nat,Zanella2022}). Here we discuss the results of our simple efficacy testing of the default sonification approach used by {\tt astronify} (a pitch-to-brightness mapping; Section~\ref{sec:sonification}), applied to synthetic light curves containing one or two transit-like signals with a range of SNRs (Figure~\ref{fig:examples}; Section~\ref{sec:mocks}). 

\subsection{Summary and expected need for sonification training}
\label{sec:training}
Our results have shown that the high SNR signals (SNR$\gtrsim$30) can be successfully identified using a sonification data format (constructed with the default approach of {\tt astronify}), by volunteer participants irrespective of previous astronomy and data analysis experience (Figure~\ref{fig:successRates}). This is a promising start and is encouraging towards the wider goals of using sonification for applications such as enhancing accessibility and live data monitoring (see discussion in Section~\ref{sec:intro}). 

In contrast, the results are not as immediately promising for using this sonification approach for medium to low SNRs, where the participants had low success rates at correctly identifying the signals in the data with sonification ($\lesssim$25\%). For the Expert group their performance with visual data representations was significantly better than using the sonifications in the medium SNR regime (SNRs=7--10). They also performed significantly better than the Non Expert group in this SNR regime with visual data, whereas the Non Expert group showed no difference between visuals and sonifications. This might not be surprising, considering that the Expert group, which contains people who have ``PhD/academic'' level experience in data analysis and astronomy, are expected to have had extensive training and experience in exploring these type of data visually. 

The comparable performance with using sonification across both expertise groups can be explained considering that they are both very unlikely to have had significant experience in using {\tt astronify}, or sonification in general, to inspect data for signals. Another hint that the Expert group responds similarly to the Non Expert group, when presented with the unfamiliar sonification format, is the Expert group's increased tendency to guess their answers when no visuals are present compared to when visuals are present (see Section~\ref{Sec:Guessing}) and dashed lines in Figure~\ref{fig:demographics}).

We conclude that, we must expect at least some level of training and experience for sonification to be used effectively in the professional astronomy setting. To what level training is important, and how to optimise this training, could be assessed with focus groups and/or similar ``success rate'' tests but comparing groups that have, versus have not, received appropriate sonification training. 

\subsection{Combining sonifications and visualisations}
\label{sec:combo}
Our results reveal no significant difference in success rates of signal identification between the data formats of plots and plots+sonifications. With the exception of the data with SNR=100 (where plots+sonifications yielded slightly higher success rates), this is true across all SNRs and across both expertise groups. There is some possible tendency for the combined data formats to systematically yield the lowest success rates in the low and medium SNR regimes; although this is not statistical significant with the current results (Figure~\ref{fig:successRates} and Figure~\ref{fig:snr710}). Our results imply that, under these testing conditions, these particular sonifications have not added any additional benefit for signal identification above and beyond using the visualisations only. 

This conclusion is consistent with the findings of \cite{DiazMerced2013} who developed upon the older sonification tool {\tt xSonify} (\citealt{Candey2006}) to sonify similar types of simulated data to that used here and performed testing with a small number of specialised participants. However, they found that adding another cue, in the form of a line sweeping across the visual display at the rate the sound was played, resulted in the sonifications actually helping the participants identify events masked by noise. \cite{DiazMerced2013} suggested that this simultaneous visual queue, congruent with the sound can help control perception of an ambiguous sound. It remains unclear whether there are preferential data to sound mapping approaches to be used when pairing up sonifications and visuals, which may even be different to the preferred mapping for a sonification only format. 

\subsection{Extending this study and further testing}
\label{sec:limitations}

A extension of this work would be to perform testing with smaller groups of people under more controlled conditions. For example, this would allow regulation of how long the participants inspected the data and ensuring the sound quality was consistent throughout (e.g., consistent use of the same playback devices). This approach could also be useful for a better understanding of the individual expertise and experience of the individual participants, which appears to be of crucial importance (Section~\ref{sec:training}). For example, musical experience may also play a roll in the level performance in signal detection using sonifications versus visualisations. Indeed, one highly qualified musician who participated (with no astronomy or data analysis experience) noted that they found the visuals distracting and preferred to close their eyes and listen only when inspecting the plots+sonifications representations (Leigh Harrison, private communication).

Testing not just the ``success'' but also the speed in which signals were identified by participants would be another step towards understanding the benefits of sonification; for example, if this approach is to be applied to live data monitoring \citep[][]{Cooke2019}. Testing different approaches for the mapping of data to sound for achieving both accuracy and speed will also be important (see Section~\ref{sec:future}). 

Our current results can not guarantee similar results for efficacy testing of more complex data (e.g., with more dimensions, uneven data sampling or less regular signals). Therefore, another extension of this work would be to simulate, and then sonify, more realistic light curves for specific applications (such as exoplanet transits or X-ray binary eclipses; e.g., \citealt{Salisbury2021,Knight2022}) with transit depths, shapes and periods that are based on real data. Furthermore, testing not just the ability for participants to use sonification to detect/count signals, but also to characterise the signals (e.g., the transit lengths or shapes) would be another interesting avenue to explore. 

\begin{figure}
\includegraphics[width=\linewidth]{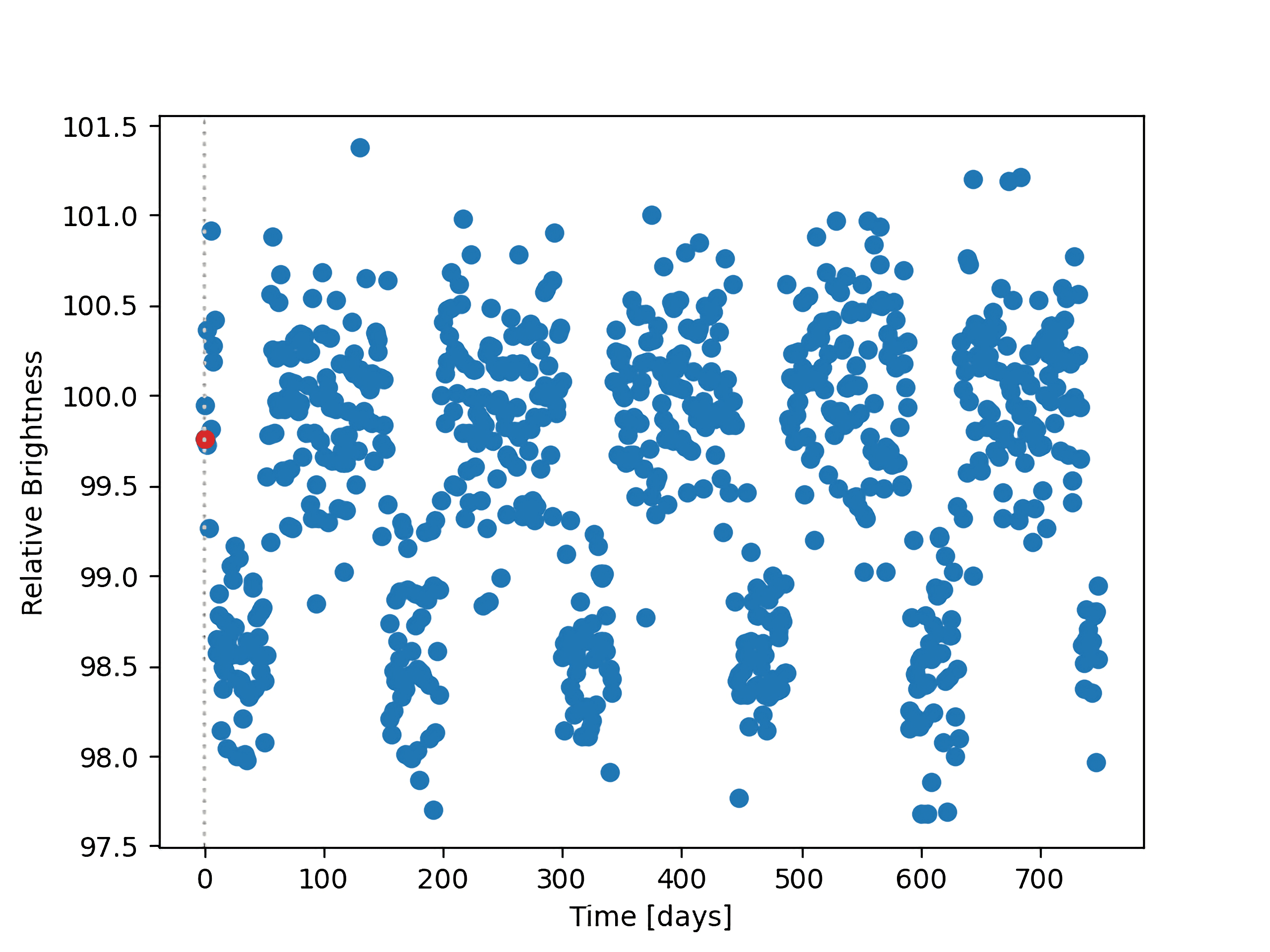}
\caption{A synthetic light curve with 6 transit-like signals, for which an animated plot+sonification has been produced using: (1) the default {\tt astronify} sonification mapping approach (link to sonification movie: \href{https://data.ncl.ac.uk/articles/dataset/Tucker-Brown_et_al_-_Astronify_Efficacy_Testing_-_Data/20936749?file=37332523}{here}) and; (2) our new filter--cut-off approach (link to sonification movie: \href{https://data.ncl.ac.uk/articles/dataset/Tucker-Brown_et_al_-_Astronify_Efficacy_Testing_-_Data/20936749?file=37332520}{here}; Section~\ref{sec:future}). In both cases the movie has been created with an animated scroll bar (grey dashed line), and a moving red data point, as a visual cue to which data points can be heard in the sonification at that time.}
\label{fig:scrollCutOff}
\end{figure}

\subsection{Future development of sonification approaches}
\label{sec:future}
The sonification approach employed through {\tt astronify} is relatively basic, although there are some parameters that can be varied; for example, the functional form of the data to pitch mapping. Here, we only tested the default approach of linear mapping. We suggest some simple additional options, and approaches, that could be implemented and tested within {\tt astronify} or other sonification tools. Firstly, adding a built-in option to produce a movie for a combined plots+sonification, with a sweeping line and highlighted data point to provide an animated visual cue of the sonification\footnote{We note that such an animation can be created using external {\tt Python} modules to {\tt astronify} and including this as a feature {\em within} {\tt astronify} is planned. This is discussed on the github pages: \url{https://github.com/spacetelescope/astronify/issues/47}.}, motivated by the discussion in Section~\ref{sec:combo}. Our example of how this might work, using a synthetic light curve containing six transit-like features, is presented in Figure~\ref{fig:scrollCutOff}. 

Secondly, allowing the user to interactively explore the sonification, scanning through the data in an exploratory manner more similar to that naturally done by the eye as it scans over a visualisation. This might include a graphical user interface, which has been included in some other sonification tools being developed within the astronomical community (e.g., {\tt StarSound} and {\tt SonoUno}\footnote{\url{http://sion.frm.utn.edu.ar/sonoUno/}}). This would have the advantage of not having to repeatedly play the sonification from the very beginning, and instead the user could focus on a region of interest. 

Finally, we propose that more complex data and more complex mappings of data to sound should be considered and tested; considering the use of timbre, volume, stereo panning etc. in addition to simply pitch. For example, we have sonified the synthetic light curve, shown in Figure~\ref{fig:scrollCutOff}, using a method that represents the data using a swept filter applied to a continuous, harmonically rich tone. Specifically, we created a musical chord and then applied a \citet{Butterworth30} low-pass filter with a 12 dB roll-off, where the cut-off frequency is mapped to the brightness of each data point with positive polarity, analogous to the pitch mapping of {\tt astronify}. The two approaches can be compared following the links in the caption of Figure~\ref{fig:scrollCutOff}. Our new filter--cut-off approach has the effect of a changing timbre or ``brightness'' of the sound when a transit is present in the data due to the changing harmonic content of the resultant sound, with dimmer fluxes yielding a duller sound. Furthermore, this sonification results in a less chaotic sound, particularly for noisy data, than occurs with the direct and discrete one-to-one mapping of individual data points to pitches. This filter cut-off approach to sonify univariate one-dimensional data has been implemented into our {\tt STRAUSS} {\tt Python} package (\citealt{STRAUSS}; also see \citealt{Harrison2022AG}) and an example code, `SonifyingData1D.ipynb' can be found on github: \url{https://github.com/james-trayford/strauss/tree/main/examples}. 

Exploring users' preferences from a range of different sonification approaches will help guide further developments. Such a study, exploring volunteer participants' preferences between different sonification approaches of magnetospheric ultra-low-frequency waves, was performed recently by \cite{Archer2022}.

Overall, we feel encouraged that the astronomical community is taking positive steps towards unlocking the huge potential of sonification, with the ultimate goal to make this a standard feature within major astronomical archives and data analysis software. We hope that this study sets the scene for more rigorous efficacy testing and evidence-based development of sonification tools, moving forward.

\section{Conclusions}
\label{sec:Conclusions}
We have performed the first large-scale user testing of {\tt astronify}, a new {\tt Python}-based sonification tool that is being developed, primarily for the sonification of astronomical data available through the MAST archive. We used the package to produce synthetic light curves containing zero, one, or two signal `transits' and then produced visual plots, sonifications, and plots+sonifications of these data using the default {\tt astronify} approach (Figure~\ref{fig:examples}). We produced data with a range of SNRs from 3--100, and performed a remote survey to assess volunteer participants' ability to identify the correct number of signals when presented with a random set of data representations. We received 192 survey responses, each of which contained answers to a random set of nine light curve data representations. Participants were grouped as astronomy and data analysis `Experts' or `Non Experts' (Figure~\ref{fig:demographics}; Section~\ref{sec:demographics}). Our main conclusions are as follows:

\begin{itemize}
    \item Participants across both expertise groups were effective at identifying high SNR signals (SNR=30 and 100) when presented with the data visually, or with a sonification, with success rates of 83-100\% in correctly counting the number of signals across the three data formats (Figure~\ref{fig:successRates}). 
    \item For low SNRs (SNR=3 and 5) participants across both groups were ineffective at identifying the signals across all three data formats, with low success rates (0--27\%) and a performance that was mostly consistent with random guessing. In this SNR regime, only the Expert group performed better than guessing and only using plots or plots+sonifications with a SNR of 5. Nonetheless, this result is driven by the Expert group's tendency to {\em guess} less frequently that there were signals in the visual data representations (compared to the Non Expert group) and their success rate remained low with these data representations including visual plots (21--24\%; Figure~\ref{fig:successRates}). 
    \item For medium SNRs (SNR=7 and 10), the Expert group performed better than random guessing when presented with the sonified only data, but with modest success rates of 26$^{+5}_{-4}$\%. However, this same Expert group performed significantly better, by a factor of $\sim$2.5, when presented with plots or plots+sonifications. The Non Expert group performed equally well to the Expert group for sonifications in this medium SNR regime (success rates of 25$_{-8}^{+10}$\%). However, unlike the Expert group, they did not show a significantly better performance with plots or plots+sonifications compared to sonifications only (Figure~\ref{fig:snr710}). 
\end{itemize}

Sonification has a significant potential to aid exploration of astronomical data, as well as to increase accessibility to those who find visual representations of data difficult, or impossible, to interpret. Our results show that the simple default sonification method within {\tt astronify} has potential utility for identifying high SNR signals in univariate light curves for users with little-to-no experience with such data represented with sound. Further testing with more realistic data sets would be the natural next step to confirm these conclusions.

However, our results imply that training and/or experience in using this particular sonification approach is likely required for this to be as useful (or more useful) than visual representations for existing experts in astronomy/data analysis. Nonetheless, this study is limited by exploring only one sonification approach and more sophisticated and developed approaches (e.g., Figure~\ref{fig:scrollCutOff}) may yield even more promising results towards a wider application of sonification in astronomy. Further participant studies testing different sonification approaches, and dedicated studies towards sonification as a route for increasing accessibility to astronomy research, are clearly warranted and are encouraged by the authors. 

\section*{Acknowledgements}
We thank the referee for their excellent comments and suggestions. We thank all of the anonymous, volunteer participants who carried out the surveys. We thank the developers of {\tt astronify} (Clara Brasseur, Scott Fleming, Jennifer Kotler, Kate Meredith) for making their code available on a open-source basis and for their consultation throughout the work. CMH acknowledges funding from an United Kingdom Research and Innovation grant (code: MR/V022830/1). During this work we made use of the software {\tt ffmpeg} (\citealt{ffmpeg}), and {\tt Python} packages {\tt matplotlib} (\citealt{matplotlib}),  {\tt astropy} (\citealt{astropy1,astropy2}) and {\tt numpy} (\citealt{numpy}).

\section*{Data Availability}
The full set of 54 data representations presented to participants in each of the three formats, a table containing all of the survey responses, a transcript of the survey text and tutorial examples and the code used for the analysis are all available here: \url{https://doi.org/10.25405/data.ncl.20936749}.



\bibliographystyle{mnras}

\begin{thebibliography}{}
\makeatletter
\relax
\def\mn@urlcharsother{\let\do\@makeother \do\$\do\&\do\#\do\^\do\_\do\%\do\~}
\def\mn@doi{\begingroup\mn@urlcharsother \@ifnextchar [ {\mn@doi@}
  {\mn@doi@[]}}
\def\mn@doi@[#1]#2{\def\@tempa{#1}\ifx\@tempa\@empty \href
  {http://dx.doi.org/#2} {doi:#2}\else \href {http://dx.doi.org/#2} {#1}\fi
  \endgroup}
\def\mn@eprint#1#2{\mn@eprint@#1:#2::\@nil}
\def\mn@eprint@arXiv#1{\href {http://arxiv.org/abs/#1} {{\tt arXiv:#1}}}
\def\mn@eprint@dblp#1{\href {http://dblp.uni-trier.de/rec/bibtex/#1.xml}
  {dblp:#1}}
\def\mn@eprint@#1:#2:#3:#4\@nil{\def\@tempa {#1}\def\@tempb {#2}\def\@tempc
  {#3}\ifx \@tempc \@empty \let \@tempc \@tempb \let \@tempb \@tempa \fi \ifx
  \@tempb \@empty \def\@tempb {arXiv}\fi \@ifundefined
  {mn@eprint@\@tempb}{\@tempb:\@tempc}{\expandafter \expandafter \csname
  mn@eprint@\@tempb\endcsname \expandafter{\@tempc}}}

\bibitem[\protect\citeauthoryear{Alexander, Gilbert, Landi, Simoni, Zurbuchen
  \& Roberts}{Alexander et~al.}{2011}]{Alexander2011}
Alexander R.,  Gilbert J.,  Landi E.,  Simoni M.,  Zurbuchen T.,   Roberts D.,
  2011, in 17th International Conference on Auditory Display (ICAD2011).

\bibitem[\protect\citeauthoryear{{Alexander}, {O'Modhrain}, {Roberts},
  {Gilbert}  \& {Zurbuchen}}{{Alexander} et~al.}{2014}]{Alexander2014}
{Alexander} R.~L.,  {O'Modhrain} S.,  {Roberts} D.~A.,  {Gilbert} J.~A.,
  {Zurbuchen} T.~H.,  2014, \mn@doi [Journal of Geophysical Research (Space
  Physics)] {10.1002/2014JA020025}, \href
  {https://ui.adsabs.harvard.edu/abs/2014JGRA..119.5259A} {119, 5259}

\bibitem[\protect\citeauthoryear{{Andreoni} \& {Cooke}}{{Andreoni} \&
  {Cooke}}{2019}]{Andreoni2019}
{Andreoni} I.,  {Cooke} J.,  2019, in {Griffin} R.~E.,  ed., ~ Vol. 339,
  Southern Horizons in Time-Domain Astronomy. pp 135--138 (\mn@eprint {arXiv}
  {1802.01100}), \mn@doi{10.1017/S1743921318002399}

\bibitem[\protect\citeauthoryear{{Archer}, {Hartinger}, {Redmon},
  {Angelopoulos}  \& {Walsh}}{{Archer} et~al.}{2018}]{Archer2018}
{Archer} M.~O.,  {Hartinger} M.~D.,  {Redmon} R.,  {Angelopoulos} V.,   {Walsh}
  B.~M.,  2018, \mn@doi [Space Weather] {10.1029/2018SW001988}, \href
  {https://ui.adsabs.harvard.edu/abs/2018SpWea..16.1753A} {16, 1753}

\bibitem[\protect\citeauthoryear{{Archer}, {Cottingham}, {Hartinger}, {Shi},
  {Coyle}, {Hill}, {Fox}  \& {Masongsong}}{{Archer} et~al.}{2022}]{Archer2022}
{Archer} M.~O.,  {Cottingham} M.,  {Hartinger} M.~D.,  {Shi} X.,  {Coyle} S.,
  {Hill} E.~D.,  {Fox} M. F.~J.,   {Masongsong} E.~V.,  2022, \mn@doi
  [Frontiers in Astronomy and Space Sciences] {10.3389/fspas.2022.877172},
  \href {https://ui.adsabs.harvard.edu/abs/2022FrASS...9.7172A} {9, 877172}

\bibitem[\protect\citeauthoryear{{Astropy Collaboration} et~al.,}{{Astropy
  Collaboration} et~al.}{2013}]{astropy1}
{Astropy Collaboration} et~al., 2013, \mn@doi [\aap]
  {10.1051/0004-6361/201322068}, \href
  {http://adsabs.harvard.edu/abs/2013A\%26A...558A..33A} {558, A33}

\bibitem[\protect\citeauthoryear{{Astropy Collaboration} et~al.,}{{Astropy
  Collaboration} et~al.}{2018}]{astropy2}
{Astropy Collaboration} et~al., 2018, \mn@doi [\aj] {10.3847/1538-3881/aabc4f},
  \href {https://ui.adsabs.harvard.edu/abs/2018AJ....156..123A} {156, 123}

\bibitem[\protect\citeauthoryear{{Bardelli}, {Ferretti}, {Ludovico}, {Presti}
  \& {Rinaldi}}{{Bardelli} et~al.}{2022}]{Bardelli2022}
{Bardelli} S.,  {Ferretti} C.,  {Ludovico} L.~A.,  {Presti} G.,   {Rinaldi} M.,
   2022, in Proceedings of "Interactive Cultural Heritage and Arts", 23rd HCI
  International Conference.  (\mn@eprint {arXiv} {2202.05539}),
  \mn@doi{10.1007/978-3-030-77411-0_12}

\bibitem[\protect\citeauthoryear{{Bieryla}, {Hyman}  \& {Davis}}{{Bieryla}
  et~al.}{2020}]{Bieryla2020}
{Bieryla} A.,  {Hyman} S.,   {Davis} D.,  2020, in American Astronomical
  Society Meeting Abstracts \#235. p. 203.04

\bibitem[\protect\citeauthoryear{{Borucki} et~al.,}{{Borucki}
  et~al.}{2010}]{Borucki2010}
{Borucki} W.~J.,  et~al., 2010, \mn@doi [Science] {10.1126/science.1185402},
  \href {https://ui.adsabs.harvard.edu/abs/2010Sci...327..977B} {327, 977}

\bibitem[\protect\citeauthoryear{Butterworth}{Butterworth}{1930}]{Butterworth30}
Butterworth S.,  1930, Experimental Wireless and the Wireless Engineer, 7

\bibitem[\protect\citeauthoryear{{Candey}, {Diaz Merced}  \&
  {Schertenleib}}{{Candey} et~al.}{2006}]{Candey2006}
{Candey} R.,  {Diaz Merced} W.,   {Schertenleib} A.,  2006, in International
  Community for Auditory Display (2006). pp 289--290

\bibitem[\protect\citeauthoryear{{Casado}, {De La Vega}, {D{\'\i}az-Merced},
  {Gandhi}  \& {Garc{\'\i}a}}{{Casado} et~al.}{2021}]{Casado2021}
{Casado} J.,  {De La Vega} G.,  {D{\'\i}az-Merced} W.,  {Gandhi} P.,
  {Garc{\'\i}a} B.,  2021, in {Ros} R.~M.,  {Garc{\'\i}a} B.,  {Gullberg}
  S.~R.,  {Mold{\'o}n} J.,   {Rojo} P.,  eds, ~ Vol. 367, Education and
  Heritage in the Era of Big Data in Astronomy. pp 120--123,
  \mn@doi{10.1017/S174392132100079X}

\bibitem[\protect\citeauthoryear{{Cooke}, {D{\'\i}az-Merced}, {Foran}, {Hannam}
   \& {Garcia}}{{Cooke} et~al.}{2019}]{Cooke2019}
{Cooke} J.,  {D{\'\i}az-Merced} W.,  {Foran} G.,  {Hannam} J.,   {Garcia} B.,
  2019, in {Griffin} R.~E.,  ed., ~ Vol. 339, Southern Horizons in Time-Domain
  Astronomy. pp 251--256, \mn@doi{10.1017/S1743921318002703}

\bibitem[\protect\citeauthoryear{Diaz-Merced}{Diaz-Merced}{2013}]{DiazMerced2013}
Diaz-Merced W.,  2013, PhD thesis, School of Computing Science, University of
  Glasgow

\bibitem[\protect\citeauthoryear{{Diaz-Merced}, {Candey}, {Mannone}, {Fields}
  \& {Rodriguez}}{{Diaz-Merced} et~al.}{2008}]{DiazMerced2008}
{Diaz-Merced} W.~L.,  {Candey} R.~M.,  {Mannone} J.~C.,  {Fields} D.,
  {Rodriguez} E.,  2008, Sun and Geosphere, \href
  {https://ui.adsabs.harvard.edu/abs/2008SunGe...3...42D} {3, 42}

\bibitem[\protect\citeauthoryear{{Garc{\'\i}a-Benito} \&
  {P{\'e}rez-Montero}}{{Garc{\'\i}a-Benito} \&
  {P{\'e}rez-Montero}}{2022}]{GarciaBenito2022}
{Garc{\'\i}a-Benito} R.,  {P{\'e}rez-Montero} E.,  2022, arXiv, \href
  {https://ui.adsabs.harvard.edu/abs/2022arXiv220512984G} {p. arXiv:2205.12984}

\bibitem[\protect\citeauthoryear{{Garcia}, {Diaz-Merced}, {Casado}  \&
  {Cancio}}{{Garcia} et~al.}{2019}]{Garcia2019}
{Garcia} B.,  {Diaz-Merced} W.,  {Casado} J.,   {Cancio} A.,  2019, in European
  Physical Journal Web of Conferences. p. 01013,
  \mn@doi{10.1051/epjconf/201920001013}

\bibitem[\protect\citeauthoryear{{Guttman}, {Gilroy}  \& {Blake}}{{Guttman}
  et~al.}{2005}]{Guttman2005}
{Guttman} S.,  {Gilroy} L.,   {Blake} R.,  2005, Psychol Sci., 16(3), 228

\bibitem[\protect\citeauthoryear{Harris et~al.,}{Harris et~al.}{2020}]{numpy}
Harris C.~R.,  et~al., 2020, \mn@doi [Nature] {10.1038/s41586-020-2649-2}, 585,
  357

\bibitem[\protect\citeauthoryear{{Harrison}, {Zanella}, {Bonne}, {Meredith}  \&
  {Misdariis}}{{Harrison} et~al.}{2022a}]{Harrison2022Nat}
{Harrison} C.,  {Zanella} A.,  {Bonne} N.,  {Meredith} K.,   {Misdariis} N.,
  2022a, \mn@doi [Nature Astronomy] {10.1038/s41550-021-01582-y}, \href
  {https://ui.adsabs.harvard.edu/abs/2022NatAs...6...22H} {6, 22}

\bibitem[\protect\citeauthoryear{{Harrison}, {Trayford}, {Harrison}  \&
  {Bonne}}{{Harrison} et~al.}{2022b}]{Harrison2022AG}
{Harrison} C.,  {Trayford} J.,  {Harrison} L.,   {Bonne} N.,  2022b, \mn@doi
  [Astronomy and Geophysics] {10.1093/astrogeo/atac027}, \href
  {https://ui.adsabs.harvard.edu/abs/2022A&G....63.2.38H} {63, 2.38}

\bibitem[\protect\citeauthoryear{Hermann, Hunt  \& Neuhoff}{Hermann
  et~al.}{2011}]{Hermann2011}
Hermann T.,  Hunt A.,   Neuhoff J.~G.,  2011, The {Sonification Handbook}.
Logos Verlag, Berlin

\bibitem[\protect\citeauthoryear{Hunter}{Hunter}{2007}]{matplotlib}
Hunter J.~D.,  2007, \mn@doi [Computing in Science \& Engineering]
  {10.1109/MCSE.2007.55}, 9, 90

\bibitem[\protect\citeauthoryear{{Knight}, {Ingram}, {Middleton}  \&
  {Drake}}{{Knight} et~al.}{2022}]{Knight2022}
{Knight} A.~H.,  {Ingram} A.,  {Middleton} M.,   {Drake} J.,  2022, \mn@doi
  [\mnras] {10.1093/mnras/stab3722}, \href
  {https://ui.adsabs.harvard.edu/abs/2022MNRAS.510.4736K} {510, 4736}

\bibitem[\protect\citeauthoryear{Kramer et~al.,}{Kramer
  et~al.}{1999}]{Kramer1999}
Kramer G.,  et~al., 1999, in The sonification report: Status of the field and
  research agenda. Report prepared for the National Science Foundation by
  members of the International Community for Auditory Display.

\bibitem[\protect\citeauthoryear{{Landi}, {Alexander}, {Gruesbeck}, {Gilbert},
  {Lepri}, {Manchester}  \& {Zurbuchen}}{{Landi} et~al.}{2012}]{Landi2012}
{Landi} E.,  {Alexander} R.~L.,  {Gruesbeck} J.~R.,  {Gilbert} J.~A.,  {Lepri}
  S.~T.,  {Manchester} W.~B.,   {Zurbuchen} T.~H.,  2012, \mn@doi [\apj]
  {10.1088/0004-637X/744/2/100}, \href
  {https://ui.adsabs.harvard.edu/abs/2012ApJ...744..100L} {744, 100}

\bibitem[\protect\citeauthoryear{{Lightkurve Collaboration}
  et~al.,}{{Lightkurve Collaboration} et~al.}{2018}]{LightKurve}
{Lightkurve Collaboration} et~al., 2018, {Lightkurve: Kepler and TESS time
  series analysis in Python} (\mn@eprint {ascl} {1812.013})

\bibitem[\protect\citeauthoryear{{Noel-Storr} \& {Willebrands}}{{Noel-Storr} \&
  {Willebrands}}{2022}]{NoelStorr2022}
{Noel-Storr} J.,  {Willebrands} M.,  2022, Nature Astronomy in press,
  arXiv:2206.13815, \mn@doi{https://doi.org/10.1038/s41550-022-01691-2}

\bibitem[\protect\citeauthoryear{{Ricker} et~al.,}{{Ricker}
  et~al.}{2015}]{Ricker2015}
{Ricker} G.~R.,  et~al., 2015, \mn@doi [Journal of Astronomical Telescopes,
  Instruments, and Systems] {10.1117/1.JATIS.1.1.014003}, \href
  {https://ui.adsabs.harvard.edu/abs/2015JATIS...1a4003R} {1, 014003}

\bibitem[\protect\citeauthoryear{{Salisbury}, {Kolb}, {Norton}  \&
  {Haswell}}{{Salisbury} et~al.}{2021}]{Salisbury2021}
{Salisbury} M.~A.,  {Kolb} U.~C.,  {Norton} A.~J.,   {Haswell} C.~A.,  2021,
  \mn@doi [\na] {10.1016/j.newast.2020.101477}, \href
  {https://ui.adsabs.harvard.edu/abs/2021NewA...8301477S} {83, 101477}

\bibitem[\protect\citeauthoryear{Sawe, Chafe  \& Treviño}{Sawe
  et~al.}{2020}]{Sawe2020}
Sawe N.,  Chafe C.,   Treviño J.,  2020, \mn@doi [Frontiers in Communication]
  {10.3389/fcomm.2020.00046}, 5

\bibitem[\protect\citeauthoryear{{Scarf}, {Gurnett}, {Kurth}  \&
  {Poynter}}{{Scarf} et~al.}{1982}]{Scarf1982}
{Scarf} F.~L.,  {Gurnett} D.~A.,  {Kurth} W.~S.,   {Poynter} R.~L.,  1982,
  \mn@doi [Science] {10.1126/science.215.4532.587}, \href
  {https://ui.adsabs.harvard.edu/abs/1982Sci...215..587S} {215, 587}

\bibitem[\protect\citeauthoryear{{Shams}}{{Shams}}{2000}]{Shams2000}
{Shams} L.,  2000, Nature, 408, 788

\bibitem[\protect\citeauthoryear{Tomar}{Tomar}{2006}]{ffmpeg}
Tomar S.,  2006, Linux Journal, 2006, 10

\bibitem[\protect\citeauthoryear{{Trayford}}{{Trayford}}{2021}]{STRAUSS}
{Trayford} J.,  2021, james-trayford/strauss: v0.1.0 Pre-release,
  \mn@doi{10.5281/zenodo.5776280}

\bibitem[\protect\citeauthoryear{{Zanella}, {Harrison}, {Lenzi}, {Cooke},
  {Damsma}  \& {Fleming}}{{Zanella} et~al.}{2022}]{Zanella2022}
{Zanella} A.,  {Harrison} C.~M.,  {Lenzi} S.,  {Cooke} J.,  {Damsma} P.,
  {Fleming} S.~W.,  2022, Nature Astronomy, in press,  arXiv:2206.13536,
  \mn@doi{https://doi.org/10.1038/s41550-022-01721-z}

\makeatother
\end{thebibliography}


\bsp	
\label{lastpage}
\end{document}